\documentstyle[prd,preprint,aps]{revtex}

\newcommand{\be}{\begin{equation}}
\newcommand{\ee}{\end{equation}}
\newcommand{\bea}{\begin{eqnarray}}
\newcommand{\eea}{\end{eqnarray}}

\begin{document}

\reversemarginpar
\tighten

\title{A  brief commentary on  black hole entropy}

\author {A.J.M. Medved}

\address{
School of Mathematics, Statistics  and Computer Science\\
Victoria University of Wellington\\
PO Box 600, Wellington, New Zealand \\
E-Mail: joey.medved@mcs.vuw.ac.nz}

\maketitle

\begin{abstract}

It is commonplace, in the literature, to find that
the Bekenstein--Hawking entropy  has  been endowed
with having an explicit  statistical interpretation. In the following essay,
we discuss why such a viewpoint warrants a certain 
degree of caution.

\end{abstract}

\section*{}

It is generally accepted as a fact that black holes are
thermodynamic objects  having, in particular, 
well-defined notions of temperature 
 and entropy.  The latter of these attributes ---   or the 
Bekenstein--Hawking entropy \cite{Bek,Haw} --- 
takes on a remarkably simple form  in Planck units: it
is just
one quarter of the horizon  cross-sectional area
or $S=A/4$.~\footnote{Note that, throughout, all fundamental constants
are set to unity and a macroscopically large black hole is assumed.}   

In light of the ``laws'' of   black hole mechanics \cite{Bard},
the ``generalized second law'' of thermodynamics \cite{Bek}
and the (almost) undisputed existence of Hawking radiation \cite{Haw},
the Bekenstein--Hawking entropy is
on rather firm ground  from a thermodynamic perspective.
What is also commonly accepted --- although on somewhat
shakier terrain --- is that this entropy should have an underlying  statistical
explanation.  Which is to say, one expects to be able to write
$S=\ln n$ for some positive integer $n>>1$. If this were indeed the case,
then the entropy would have a statistical meaning in the usual sense;
namely, as a logarithmic measure of the number
 ($n$) of fundamental microstates. No proponent of
this viewpoint can be quite sure as to  what exact
 form these constituents might take,
but the consensus feeling is that the microstates should  be   traceable to
the (currently enigmatic) theory of quantum gravity.

Speaking of the quantum aspects of gravity, there
are strong reasons to believe that black holes are,
themselves, fundamentally quantum entities.  
If this is an accurate assessment, 
it naturally 
follows that any associated
classical observable should have a quantum-operative counterpart.
In particular, the possibility of a quantum  spectrum for
the horizon area of a black hole has received 
considerable  attention; beginning with the intuitive arguments
of Bekenstein \cite{Bek2} and cumulating with the rigorous assertions 
of loop quantum gravity \cite{Ashk}. 
The plot thickens considerably when one combines 
this concept of a quantized area with the above idea about statistics.
For instance, let us suppose that, between quantum area levels,
 there is some {\it minimal} spacing    $\delta A$.  
Let us further suppose
that, by virtue of the quantum-uncertainty principle,
this minimal gap is of Planck order or
$\delta A \sim {\cal O}[1]$. ~\footnote{The 
loop quantum gravity 
supporters will undoubtedly point out
that the spacing is expected to become decreasingly and
exponentially  small 
as the black hole becomes macroscopically large \cite{Rove}. Nonetheless,
there are still valid arguments --- even in the context of
this framework  ---  that support the idea of 
a  minimal spacing, at least under certain circumstances 
\cite{Poly,Alex,Smol}.}
Then, given that the entropy has a statistical basis,
the relation  $\delta A= 4\ln k$ (for some positive integer
$k<<n$) immediately follows \cite{Muk}. It is notable that such 
an idea has inspired the  ``Hod conjecture'' \cite{Hod};
which professes to provide a quantum-gravitational meaning
to the $\ln 3$ which notoriously turns up
in the quasinormal-mode spectrum of Schwarzschild-like black
holes \cite{Motl}.

In apparent contradiction with the above discussion,
not all derivations of the black hole area spectrum substantiate
a logarithmic spacing. Here, we will take note of
three such treatments: \\
{\it (i)} a  method based on  reducing the black hole
phase space to a pair of observables (the mass  and its conjugate) 
and then elevating these to quantum operators \cite{Barv}, \\
{\it (ii) } a method based on identifying the exponent
of  the gravitational action as a quantum amplitude and then enforcing 
observer independence  on its value at a horizon \cite{Pady}, \\
{\it (iii)} a method based on applying non-commutative quantum theory
and then identifying the area with a number operator  \cite{Rome}
(also see \cite{Dolan}). \\
What is interesting is that all three of these treatments
--- although {\it completely independent} ---
predict an area spacing of precisely $\delta A =8\pi$.

So how do we rationalize the fact that $8\pi$ obviously fails
to comply with the statistical prediction of $4\ln k$ ?
The easy answer is that the three methods are
simply wrong; after all, each of these studies does employ
(at least) one conjectural input at some point in their respective 
calculations. Nonetheless, three of a kind is a pretty strong
``hand'', so we should at least ponder
the possibility that the $8\pi$ prediction is
correct. 

Supposing that  $8\pi$ is the quantum area gap, we then have an
immediate conflict with the statistical interpretation of the black
hole  entropy or $S=\ln n$.
Which is to say, it could no longer be accurately claimed that 
this entropy
counts the number of black hole microstates. 
But should  such a conflict be regarded as a fundamental problem? 
Not according to various researchers
({\it e.g.}, \cite{Jaco,Sork,Smolin}) who have stressed that the 
Bekenstein--Hawking entropy does not necessarily reflect the internal states
of a  black hole. 
Of particular interest, Marolf
has very recently made this point \cite{Maro} on the basis 
of a calculation that demonstrates the entropy flux across
a black hole horizon as being an observer-dependent quantity
\cite{Maro2}.
To elaborate,  if a free-falling observer  
attributes the usual (statistically counted) entropy  for an 
inward falling object,
then a ``fiducial'' ({\it i.e.}, 
external and stationary) observer will attribute a much different 
entropy for the very  same object.  Typically, the  fiducially
measured value is much smaller and, perhaps strangely,
does not even have an explicit dependence on the number of microstates 
(up to lower-order corrections). In this regard, it may
be significant that  methods {\it (i)}--{\it (iii)} all 
compute the area spectrum (at least implicitly) from the perspective
of a fiducial observer.  Evidently, this type of observer has
no direct knowledge of the microstates  associated
with objects entering the black hole  nor,  presumably,
the microstates of the  black hole itself.
It should  then   not  really be much of a surprise
if this external observer deduces a black hole entropy
(by way of the quantum area spectrum) that fails to comply with 
statistical notions.

But, if the Bekenstein--Hawking entropy is 
indeed an observer-dependent construct,
then  is there still an intrinsic entropy associated with
the black hole?  Let us consider one possible answer: It has recently 
been suggested that {\it every}
spacetime two-surface has such an intrinsic entropy and
the black hole entropy (or, actually,  any entropy associated with
a horizon) is just a  observational-based manifestation
of this deeper phenomenon \cite{Make}.
Of course, this idea is conjectural and there is a myriad of other 
possibilities to consider, but it does illustrate the following point:
Any prospective theory of quantum gravity
may have a significantly greater challenge than just reproducing
the Bekenstein--Hawking entropy (which is often viewed as
a critical ``litmus test''). A candidate theory
 may also have to provide the link between
the fundamental level and the black hole  entropy as an emergent 
semi-classical concept.

But the real moral of our story is that some
caution should be used before assigning the black hole
entropy  a strict statistical meaning. (As has, for example,
become popular in studies that consider the highly
damped spectra of black hole quasinormal modes.) 
This is not to say that such an assignment need be wrong
(for instance,  background-independent theories
seem rather robust against 
observer-related notions), but rather that some  clarification
is still needed as to what one exactly means by ``black hole entropy''
--- or  {\it any} entropy for that matter.

\section*{Acknowledgments} 

Research is supported  by
the Marsden Fund (c/o the  New Zealand Royal Society) 
and by the University Research  Fund (c/o Victoria University).
The author thanks Magrietha Medved for proof reading and
Alex Nielsen for pointing out an error in the first
version.

\end{document}